\documentstyle[prd,aps,psfig]{revtex}

\def\simlt{\stackrel{<}{{}_\sim}}  \def\simgt{\stackrel{>}{{}_\sim}}
\newcommand\be{\begin{equation}}  \newcommand\ee{\end{equation}}
\newcommand\bea{\begin{eqnarray}}  \newcommand\eea{\end{eqnarray}}
\newcommand\ba{\begin{array}}  \newcommand\ea{\end{array}}
\begin{document}
\draft \input epsf  \def\la{\mathrel{\mathpalette\fun <}}
\def\ga{\mathrel{\mathpalette\fun >}}
\def\fun#1#2{\lower3.6st\vbox{\baselineskip0st\lineskip.9st
\ialign{$\mathsurround=0pt#1\hfill##\hfil$\crcr#2\crcr\sim\crcr}}}

\twocolumn[\hsize\textwidth\columnwidth\hsize\csname
@twocolumnfalse\endcsname

\title{Neutrinos and Gauge Unification} \author{J.A. Casas$^{(1, 3)}$,
J.R. Espinosa$^{(2,3)}$, A. Ibarra$^{(1)}$  and I. Navarro$^{(1)}$}
\address{\phantom{ll}}
\address{$^{(1)}${\it IEM, CSIC, Serrano 123, 28006 Madrid, Spain}}
\address{$^{(2)}${\it IMAFF, CSIC, Serrano 113 bis, 28006 Madrid, Spain}}
\address{$^{(3)}${\it IFT, C-XVI, Univ. Aut\'onoma de Madrid, 28049
Madrid, Spain}}
\date{\today}  \maketitle
\begin{abstract}
The approximate unification of
gauge couplings is the best indirect evidence for low-energy
supersymmetry, although it is not perfect in its simplest
realizations. Given the experimental evidence for small
non-zero neutrino masses, it is plausible to extend the
MSSM with three right-handed neutrino chiral multiplets, 
with large Majorana masses below the unification scale, so that a see-saw
mechanism can be implemented.  
In this extended MSSM, the unification prediction for
the strong gauge coupling constant at $M_Z$ can be lowered by up to $\sim
5\%$, bringing it closer to the experimental value at $1\sigma$,
therefore improving significantly the accuracy of gauge coupling
unification. 
\end{abstract}
\pacs{PACS: 14.60.St, 11.30.Pb, 12.10.Kt~
IFT-UAM/CSIC-00-15~ IEM-FT-203/00~ {\sf hep-ph/0004166}}  \vskip2pc]

Gauge coupling unification is an important test for candidates to
fundamental theories beyond the Standard Model (SM). On the one hand,
gauge unification is a prediction of most of such theories. In
particular, GUT models predict the unification of the gauge couplings
at a scale $\sim M_X$, corresponding to the breakdown of the unifying
group.  Also, most of the string constructions predict unification of
all the gauge couplings, including gravity, even in the absence of a
GUT group. On the other hand, due to the present high-precision
knowledge of the gauge couplings at low energy, one can extrapolate
them to high energy through the corresponding renormalization group
equations (RGEs) and verify whether unification takes place or not.

Of course the evolution of the couplings (as described by the RGEs)
depends on the type of theory under study. In particular, unification can
depend on whether it is supersymmetric or
not, whether it has the minimal matter content or not,  whether it has
(large) extra dimensions or not, etc. Consequently, the requirement of 
gauge coupling unification offers the possibility of testing
these theories at very high scales.  In this respect, it has been known
for a long
time that non-supersymmetric gauge unification (with minimal matter
content) does not work. In contrast, supersymmetric gauge
unification takes place with remarkable accuracy at $M_X\sim 2\times
10^{16}$ GeV. This fact was confirmed in 1990 with the precise LEP
determinations of the gauge couplings. The unification occurs in such
a beautiful way that it is hard to believe that it is a mere
coincidence, and it is currently considered as a strong hint in favor 
of supersymmetry. However, there are some problems with the current
status of this issue. First, the
conventional prediction of (weakly coupled heterotic) superstrings is
that the gauge couplings (including gravity) are unified at the string
scale $M_{string}\sim 5\times  10^{17}$ GeV, which is one order of
magnitude too high compared with $M_X$. During
the last years  many new string (or string-inspired) scenarios have
appeared addressing this problem: strongly
coupled
strings, large extra dimensions, etc.  Another problem is that the
ordinary supersymmetric unification, however beautiful, is not
perfect in the simplest scenarios. A convenient way to express the
discrepancy is to start with
the $SU(2)\times U(1)_Y$  low-energy couplings $\alpha_1(M_Z)$ and
$\alpha_2(M_Z)$ to obtain both the unification scale and gauge
coupling, $M_X$ and $\alpha(M_X)$. Then, running from $M_X$ back to
$M_Z$, one gets a prediction for $\alpha_3(M_Z)$, to be compared to
the experimental value.  

Suppose then that, following for example ref.~\cite{alphath} one
has a MSSM spectrum as obtained from a minimal supergravity model.
Starting with the experimental input (in the $\overline{\rm MS}$
renormalization scheme, before conversion to the $\overline{\rm DR}$
scheme)
\begin{eqnarray}
\label{eweak}
\widehat{\alpha}_1^{-1}(M_Z)&=&58.98\pm 0.04,\nonumber\\
\widehat{\alpha}_2^{-1}(M_Z)&=&29.57\pm 0.03,\nonumber\\
M_Z&=&91.197\pm 0.007,
\end{eqnarray}
one integrates numerically two-loop RGEs,
including also the supersymmetric thresholds until one finally obtains
\cite{alphath} a prediction
$\alpha_3(M_Z)\simeq 0.13$,
to be compared to the experimental value~\cite{alphaexp}
\begin{equation}
\label{strong}
\widehat{\alpha}_3(M_Z)=0.119\pm 0.004\ ,
\end{equation}
which represents a $3\sigma$ discrepancy [other evaluations give a 
smaller experimental uncertainty in (\ref{strong}), thus leading to a 
stronger discrepancy. We choose to be rather conservative at this point].  
Several comments are in
order here. The effect of supersymmetric thresholds introduces a
dependence on the type of MSSM spectrum considered. For instance, the
larger the supersymmetric masses, the lower $\alpha_3(M_Z)$. Allowing
the squark masses to be up to 1 TeV, one gets $\alpha_3(M_Z)\simgt
0.127$. On the other hand, working in the context of  gauge-mediated
SUSY breaking, which implies the presence of extra (messenger) fields
and thus a departure from the minimal matter content, one can get
slightly lower values, $\alpha_3(M_Z)\simgt 0.125$. This alleviates
the disagreement, but it does not solve it. 
One would need a negative correction between$-3\%$  and
$-9\%$ in $\alpha_3(M_Z)$ [or a correction three times smaller in
$\alpha_3(M_X)$] in order to reconcile prediction and experiment.  It has
been argued that
the origin of the discrepancy could be  high-energy threshold
corrections. These may be of stringy or GUT origin and are
model-dependent. {\it E.g.} in a GUT scenario these corrections are due
to the appearance at the GUT scale of  incomplete $SU(5)$ multiplets,
most notably the color triplet Higgs bosons.  Demanding a
threshold correction of the correct magnitude  and sign severely
restricts the models, {\it e.g.} minimal $SU(5)$ is incompatible
with such a requirement \cite{alphath}.

In the present paper we explore the impact of massive neutrinos in the
unification of gauge couplings. Observations of the flux of atmospheric
neutrinos by SuperKamiokande~\cite{atm} provide strong evidence for
neutrino oscillations, which in turn imply that (at least two species of)
neutrinos must be massive. Preliminary results from the long-baseline
experiment K2K tend to confirm this evidence \cite{K2K}. Additional
support to non-zero neutrino masses  is given by the need of neutrino
oscillations to explain the solar neutrino flux deficit~\cite{solar}.  The
simplest and most beautiful
mechanism to account for small neutrino masses  in a natural way is
probably the see-saw mechanism \cite{seesaw}.  Its supersymmetric version
has superpotential
\bea
\label{superp}
W=W_{MSSM} - \frac{1}{2}\nu_R^c{\cal M}\nu_R^c +\nu_R^c {\bf Y_\nu}
L\cdot H_2,  \eea
where $W_{MSSM}$ is the superpotential of the MSSM. The extra terms
involve, beside the usual three lepton doublets, $L$, three additional
neutrino chiral fields $\nu_R$ singlets under the SM group 
(generation indices are suppressed).  ${\bf Y_\nu}$ is the matrix of
neutrino Yukawa couplings and $H_2$ is the hypercharge $+1/2$ Higgs
doublet. The Dirac mass matrix is ${\bf m_D}={\bf Y_\nu}v
\sin\beta$.  Finally, ${\cal M}$ is a $3\times 3$ Majorana mass matrix
which does not break the SM gauge symmetry. It is natural to assume
that the overall scale of ${\cal M}$, which we will denote  by $M$, is
much larger than the electroweak scale or any soft mass. Below $M$ the
heavy neutrino fields can be integrated out, giving rise to an
effective  mass term for the left-handed neutrinos,
$-\frac{1}{2}\nu^T{\cal M}_\nu \nu$, with ${\cal M}_\nu= {\bf m_D}^T
{\cal M}^{-1} {\bf m_D}$, suppressed with respect to the typical
fermion masses by the  inverse power of the large scale $M$. In fact,
large values of the neutrino Yukawa couplings are perfectly consistent
with tiny neutrino masses for values of $M$ sufficiently close to~$M_X$.

The influence of heavy right-handed neutrinos on the unification of
gauge couplings is due to the fact that above $M$, the
Yukawa couplings ${\bf Y_\nu}$ of eq.(\ref{superp}) affect the RGEs of the
gauge
couplings at two-loop order, in a similar way as the top-Yukawa coupling 
does in the MSSM. Since ${\bf Y_\nu}$ can
be sizeable, one expects a non-trivial impact in gauge unification.
The (two-loop) RGEs at scales $Q$ between $M$
and $M_X$, for $g_i$, ${\bf Y_t}$ and ${\bf
Y_\nu}$ read [with $g_1$ normalized as in $SU(5)$]
\bea  
\frac{d g_i^{-2}}{dt}&=&-2\kappa \left[b_i+\kappa\left( 
b_{ij}g_j^2 -a_{i\alpha} {\rm tr}{\bf H}_\alpha\right)\right]
\label{RGEgi}\\
\frac{d {\bf Y_t}}{dt}&=&\kappa {\bf Y_t}\left(
-c_{i}g_i^2 +3{\bf H_t} + {\rm T}
\right)\nonumber\\
&+&\kappa^2{\bf Y_t} \left[
\left(c_ib_i+\frac{c_i^2}{2}\right)g_i^4+\frac{136}{45}g_1^2g_3^2+
8g_2^2g_3^2\right.\nonumber\\
&+&g_1^2g_2^2+\left(\frac{2}{5}g_1^2+6g_2^2\right){\bf H_t}
+\left(\frac{4}{5}g_1^2+16g_3^2\right){\rm Tr}{\bf H_t}
\nonumber\\
&-&\left.9{\rm Tr}{\bf H_t^2}-4 {\bf H_t}^2
-3 {\bf H_t} {\rm T}
-9{\rm Tr}{\bf H^2_t}\frac{}{}\right]
\label{RGEYt}\\
\frac{d {\bf Y_\nu}}{dt}&=&\kappa {\bf Y_\nu}\left(
-c'_{i}g_i^2 +3{\bf H}_\nu + {\rm T}
\right)\nonumber\\
&+&\kappa^2{\bf Y_\nu} \left[
\left(c_i'b_i+\frac{c_i'{}^2}{2}\right)g_i^4+\frac{9}{5}g_1^2g_2^2
\right.\nonumber\\
&+&6\left(\frac{1}{5}g_1^2+g_2^2\right){\bf H}_\nu
+\left(\frac{4}{5}g_1^2+16g_3^2\right){\rm Tr}{\bf H_t}
\nonumber\\
&-&\left.3{\rm Tr}{\bf H}_\nu^2-4 {\bf H}_\nu^2
-3 {\bf H}_\nu {\rm T}
-9{\rm Tr}{\bf H^2_t}\frac{}{}\right]
\label{RGEYnu}
\eea
(summation over repeated indices is understood) with $t=\ln Q$,
$\kappa=1/(16\pi^2)$ and
\be
{\bf H}_\alpha\equiv{\bf Y}_\alpha^\dagger {\bf Y}_\alpha,\qquad 
{\rm T}\equiv {\rm
Tr}(3{\bf H_t}+{\bf H}_\nu),
\ee
with $\alpha={\bf t},\nu$.
Here ${\bf Y_t}$ represents the Yukawa matrix of the $u$-type quarks
(dominated by the top Yukawa coupling) and the numerical coefficients
are given by
\bea  b_i&=&(33/5, 1, -3),\ \  c_i=(13/15, 3, 16/3), \nonumber\\
c_i'&=&(3/5, 3, 0),\qquad \ a_{i \alpha}= \left(\begin{array}{cc} 
26/5 & 6/5 \\ 
 6   &  2  \\ 
 4   &  0
\end{array}\right),  \nonumber\\
b_{ij}&=&
\left(\begin{array}{ccc} 
199/25 & 27/5 & 88/5 \\
 9/5   &  25  &  24  \\ 
11/5   &   9  &  14
\end{array}\right). 
\label{coeff}
\eea
It is clear from eq.(\ref{RGEgi}) that the presence of neutrinos will only
affect the RGEs of $g_1$ and $g_2$. The magnitude of the final correction
will depend on the numerical values of ${\bf Y_\nu}$ and $M$. 
These are not independent quantities, since the final
neutrino masses  ${\cal M}_\nu= v^2 \sin^2 \beta {\bf Y_\nu}^T {\cal M}^{-1}
{\bf Y_\nu}$ (appropriately ran down to the electroweak scale
\cite{RGmnu}) must be
consistent with observations. Atmospheric neutrino data only allow to
determine one neutrino mass splitting, $\Delta m^2_{\nu}\ \sim\
10^{-3}\ {\rm eV}^2$. On the other hand, 'standard' explanations of solar
neutrino flux deficits require neutrino oscillations with  a much smaller
mass splitting, corresponding to a different pair of neutrinos.   In
addition, there are upper bounds on neutrino masses, coming {\it e.g.} 
from the non-observation of neutrinoless double $\beta$-decay and other
experiments.  All this information implies that there are three
possible types of neutrino spectrum: Hierarchical, $m_1^2< m_2^2\ll
m_3^2$; Intermediate,  $m_1^2\sim m_2^2\gg m_3^2$; and Degenerate,
$m_1^2\sim m_2^2\sim m_3^2$.  In the hierarchical case, the larger
neutrino mass should be  ${\cal O}(10^{-1}-10^{-2})\ {\rm eV}$, while
in the degenerate case cosmological observations require
$m_\nu\simlt{\cal O}(2)\ {\rm eV}$.  This means, that once a
particular scenario for the neutrino spectrum  is chosen, ${\bf
Y_\nu}$ and $M$ are not independent quantities any more.

The presence of right-handed neutrinos above $M$ induces a 
small change both on the scale of unification and the value of the
unified gauge coupling. This in turn affects the prediction of $\alpha_3(M_Z)$
derived from the assumption of perfect unification. Given that the changes
with respect to the MSSM case can be treated as a small perturbation, it
is simple to make an analytical estimate from eqs.(\ref{RGEgi}) and 
(\ref{RGEYnu}) to get:
\be
\label{correction0}
\frac{\Delta \alpha_3(M_Z)}{\alpha_3(M_Z)^2} \simeq
-\frac{9}{7\pi}\frac{1}{16\pi^2}
<{\rm Tr} {\bf Y_\nu^\dagger Y_\nu} >\ln\frac{M_X}{M}.
\ee
In this result, $<{\rm Tr} {\bf Y_\nu^\dagger Y_\nu} >$ is an average
value
in the interval from $M$ to $M_X$:
\be
<{\rm Tr} {\bf Y_\nu^\dagger Y_\nu} >\ln\frac{M_X}{M}=\int_{Q=M}^{M_X}
{\rm Tr} {\bf Y_\nu^\dagger Y_\nu}\ d\ln Q.
\label{average}
\ee
Two important implications of Eq.~(\ref{correction0}) are, first, that
neutrino corrections always make the predicted $\alpha_3(M_Z)$ smaller, as
required to improve the agreement with the experimental value, and
second, 
that the effect will be important if the neutrino Yukawa couplings ${\bf
Y_\nu}$ are sizable, which is a natural assumption, as we saw.

We can now estimate the average in Eq.~(\ref{average}) considering
that the evolution of ${\bf Y_\nu}(t)$ is well described by its 1-loop RGE
neglecting all couplings different from  ${\bf Y_\nu}$. This
approximation is justified if ${\bf Y_\nu}$ is large (the case of
interest). The final result will depend on how many neutrino Yukawa
couplings (or more precisely, eigenvalues of
${\bf Y_\nu}^\dagger {\bf Y_\nu}$) are large. With only
one large
neutrino coupling (this corresponds naturally to a hierarchical neutrino
spectrum) it is straightforward to evaluate the integral in
(\ref{average}) in the approximation just described to get
\bea
\label{correction}
\Delta \alpha_3(M_Z) \simeq - \frac{3N}{28\pi}\alpha_3(M_Z)^2\log\left[
\frac{Y_\nu(M_X)^2}{Y_\nu(M)^2}\right],  
\eea
with $N=1$ counting the number of large neutrino Yukawa couplings.
Of course the effect on $\alpha_3(M_Z)$ is larger if all three neutrino
Yukawa couplings are large (this corresponds naturally to a degenerate
neutrino spectrum, but not necessarily: a hierarchy in the spectrum can 
be induced also in this case by
flavor-dependent Majorana masses). In this $N=3$ case, the result for
$\Delta \alpha_3(M_Z)$ slightly depends on the kind of texture for ${\bf
Y_\nu}$. A good estimate is given by simply taking ${\bf Y_\nu}= Y_\nu
{\bf 1}$ and in that case, Eq.~(\ref{correction}) also holds, simply
setting $N=3$. 

The magnitude of the correction (\ref{correction}) depends on the
values of  $Y_\nu$ at $M_X$ and $M$, but the latter is fixed in order
to fit  the physical neutrino mass, as explained above. Hence, the
correction is just a function of a unique parameter, 
$Y_\nu(M_X)$. [It is also a function of the neutrino
mass, $m_\nu$, but changes on $m_\nu$ just imply a
corresponding modification of $M$ and since the evolution of $Y_\nu$
occurs mainly near $M_X$ the final correction on
$\alpha_3(M_Z)$, given by  eq.(\ref{correction}), is quite insensitive 
to $m_\nu$.]  The most
favorable case occurs when $Y_\nu(M_X)$ is large, near a Landau
pole. A representative case is  $Y_\nu(M_X)\sim 8$. Then $Y_\nu(M)\sim
2$ and we get $\Delta \alpha_3(M_Z) \sim -1.4\%$ for the case $N=1$
 and $\Delta \alpha_3(M_Z) \sim -4.1\%$ for $N=3$.  It is clear that in
the first case this correction is not sufficient to reconcile
$\alpha_3(M_Z)$ with the experimental value, but it can be enough in the
second case. These estimates agree well with our full numerical results,
obtained by numerical integration of the RGEs (\ref{RGEgi}-\ref{RGEYt}). Now
the two-loop contributions soften the RGE for ${\bf Y_\nu}$, which implies
that ${\bf Y_\nu}$ decreases more slowly when the energy scale goes down.
This increases the Yukawa coupling average in eq.(\ref{correction0}), and
thus the (negative) shift of the low-energy strong coupling, $\Delta
\alpha_3(M_Z)$. 
\begin{figure}[hbt]
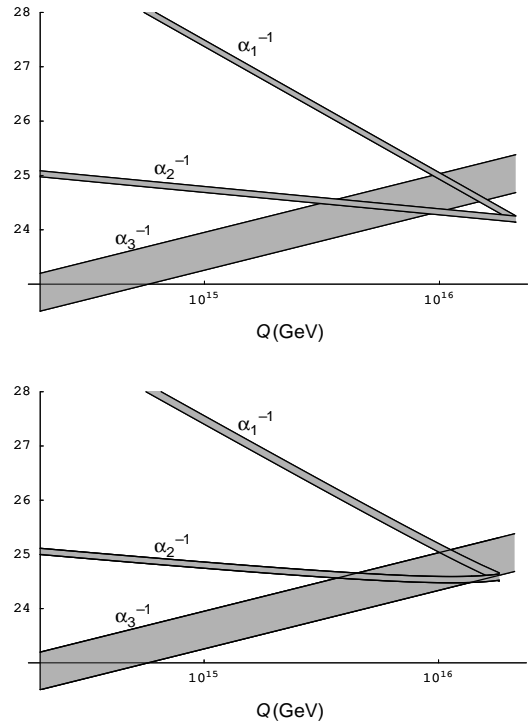

\centerline{\vbox{
\psfig{figure=unif1.ps,height=5.cm,width=7.cm,%
bbllx=8.cm,bblly=4.cm,bburx=27.cm,bbury=18.cm}
\psfig{figure=unif2.ps,height=5.cm,width=7.cm,%
bbllx=8.cm,bblly=4.cm,bburx=27.cm,bbury=18.cm}}}
\caption{\footnotesize Unification of the gauge couplings $\alpha_i^{-1}$
for a typical MSSM scenario, without massive
neutrinos (upper plot) and with them (lower plot). }
\end{figure}

An important question at this point is how large can  $Y_\nu(M_X)$ be
without jeopardizing the perturbative expansion, and hence the reliability
of the results. This has been studied in great detail in ref.\cite{QIFP} for
the case of the top Yukawa coupling. Since, in the large coupling
regime relevant for this question, the RGEs for
the top and neutrino Yukawa couplings (in the $N=3$ case) are identical, the
results of ref.~\cite{QIFP} are applicable also to the present case. 
The conclusions of the authors of ref.~\cite{QIFP}, using Pad\'e-Borel
resummations of the 4-loop beta-function, were that the qualitative
behaviour of the running is well described by the one-loop approximation and
can be further improved by Pad\'e-Borel approximants which are reliable
for values of the Yukawa coupling up to $\simlt 8$ (For
the $N=1$ case this is a conservative estimate, since in that case the
beta-function for $Y_\nu$ is softer than that of $Y_t$.). We have used a
$[1,1]$ Pad\'e-Borel resummation of the ${\bf Y_\nu}$ beta-function to
obtain our most reliable numerical results. As the beta-function in this
approximation is smaller than at one-loop, the final 
result for the average in (\ref{correction0}) turns out to be a bit larger
than our analytical estimates [for a fixed value of $Y_\nu(M_X)$]. The
negative shift on
$\alpha_3(M_Z)$ goes then from 4\% to about 5\%. This fact,
together with the uncertainties in  $\alpha_1(M_Z)$, $\alpha_2(M_Z)$ and the
top mass, imply that the usual MSSM scenarios, which yielded an
unsuccessful prediction $\alpha_3(M_Z)\simeq 0.13$, are now rescued within
$1\sigma$ confidence level. [The agreement can be perfect in
other MSSM scenarios, which predict slightly lower values of
$\alpha_3(M_Z)$, even if the error in (\ref{strong}) becomes smaller.] 

This is illustrated in fig.~1 for a typical MSSM scenario 
The first plot shows how the gauge couplings fail to unify in the ordinary MSSM
in the absence of massive neutrinos (the requirement of unification would 
imply $\alpha_3(M_Z) = 0.13$ in the case plotted).
The second plot corresponds to
the MSSM extended with neutrinos getting mass via a see-saw mechanism.
More precisely, the plot corresponds to a scenario with degenerate
neutrinos of mass $m_\nu=2$ eV and  $Y_\nu(M_X)=8$. It is apparent
how the $\alpha_1$ and $\alpha_2$ runnings are modified in a suitable
way to get gauge unification.

Besides the logarithmic effect on unification we have described, the
presence of heavy right-handed neutrinos affects the running of the gauge
couplings also through finite two-loop threshold effects at the Majorana
scale $M$. However, these will
depend on the neutrino Yukawa coupling $Y_\nu(M)$, which in the case of
interest is much smaller than $Y_\nu(M_X)$, so that it is safe to neglect
these matching effects.

Finally, it is interesting to point out that effects similar to the ones
we have found are expected in the next-to-minimal supersymmetric standard
model (NMSSM), i.e. the MSSM extended with a singlet chiral multiplet $S$. The
superpotential of the NMSSM does not contain a mass term for the Higgs
multiplets (the $\mu$-term). However, an effective $\mu$ term of the
correct order of magnitude is generated dynamically by
$\Delta W = Y_s S H_1\cdot H_2$,
when $S$ takes a vacuum expectation value. This solves in fact the $\mu$
problem of the MSSM and is one of the main virtues of the NMSSM
(although this model has its own drawbacks). The influence of the new
Yukawa coupling $Y_s$ on the running of the gauge
couplings is
also a two-loop effect, of exactly the same form as in (\ref{RGEgi}) with
$Y_s^2$ instead of ${\rm Tr}{\bf H_\nu}$ (the coefficients $a_{\alpha i}$
 are exactly the same for $\alpha=\nu$ and $\alpha=s$). 
The final impact on $\alpha_3(M_Z)$ is therefore given by
a formula like (\ref{correction0}) with ${\rm Tr}{\bf Y_\nu^\dagger
Y_\nu}$$\rightarrow$$Y_s^2$ and $M$ replaced by the mass $M_S$ of the
singlet $S$, which is close to the electroweak scale (this makes the
logarithm much larger). Numerically, we find $\Delta\alpha_3(M_Z)\simeq
-3\%$ for $M_S=1$ TeV and $Y_s(M_X)=8$. Similar effects can occur in other
extensions of the MSSM with additional Yukawa couplings, as has been shown  
for models with $R$-parity violating couplings in ref.~\cite{herbi}.

In conclusion, we have examined the impact of heavy see-saw neutrinos
(plausible in view of the growing experimental evidence  in favor of
non-zero neutrino masses) on the unification of gauge couplings, in
particular as reflected in the unification prediction for the strong gauge
coupling constant at the electroweak scale. We find
that the effect is small, but is always of the right sign and
can be of the right magnitude to bring the too high MSSM prediction for
$\alpha_3(M_Z)$ down
to values within $1\sigma$ of the experimental value. Given that adding
three heavy right-handed neutrinos is not an {\em ad-hoc} extension of the
MSSM but on the contrary is well motivated by experiment and theory alike,
this result is welcome and noticeable.
This effect should be taken into account, even in models with
sizeable stringy or GUT high-energy threshold corrections. 
For example, models that have been discarded for not giving the appropriate
threshold corrections (e.g. minimal $SU(5)$, \cite{alphath}), 
can be now perfectly consistent.

{\bf Acknowledgments} We thank H.~Dreiner for useful correspondence. 
We also thank A. Delgado for very useful discussions. A.~I.
thanks the Comunidad de Madrid (Spain) for a pre-doctoral grant.

\end{document}